\documentclass[prd,aps,twocolumn,epsfig]{revtex4}
\evensidemargin=\oddsidemargin
\usepackage{graphicx}
\usepackage{bm}
\begin{document}
\renewcommand{\baselinestretch}{1.2}

\def\del{\partial }

\begin{flushleft}
TUHEP-TH-06153\\
hep-ph/0609091
\end{flushleft}

\null\vspace{-1cm}
\title{Hadronic transition $\bm{\chi_{c1}\to\eta_c\pi\pi}$ at
the Beijing Spectrometer BES\\ and the Cornell CLEO-c}
\bigskip
\author{Qin Lu}
\address{Center for High Energy Physics, Tsinghua University, Beijing 100084, China}
\author{Yu-Ping Kuang}
\address{ CCAST (World Laboratory), P.O. Box 8730, Beijing 100080, China;\\
and Center for High Energy Physics and Department of Physics,
Tsinghua University, Beijing 100084, China\footnote{Mailing address}}

\bigskip

\begin{abstract}

Hadronic transitions of the $\chi_{cj}$ states have not been studied yet. 
We calculate the rate of the hadronic transition $\chi_{c1}\to\eta_c\pi\pi$ in the framework 
of QCD multipole expansion. We show that this process can be studied experimentally at the upgraded 
Beijing Spectrometer BES III and the Cornell CLEO-c.

\null\noindent PACS number(s): 13.20.Gd, 13.30.Eg, 14.40.Cs
\end{abstract}
\maketitle

\smallskip


It has been shown that the theory of hadronic transitions based on QCD multipole expansion 
\cite{Yan,KY81,KTY,Kuang02} can make quite successful predictions for many hadronic transition rates 
in the $c\bar{c}$ and $b\bar{b}$ systems \cite{Kuang06,brambilla}. So far, hadronic transitions of the 
$\chi_{cj}$ states have not been studied yet. The observed $\chi_{cj}$ decays are mainly hadronic 
decays (decaying into light hadrons), and the hadronic widths of the three $\chi_{cj}$
states are rather different. From the point of view of perturbative QCD, the hadronic decay rates of 
$\chi_{c0}$ and $\chi_{c2}$ are proportional to $\alpha_s^2$, while that of $\chi_{c1}$ is 
proportional to $\alpha_s^3$ \cite{hadronic}. Hence $\chi_{c1}$ has the smallest hadronic decay rate, 
and is thus the most interesting one among the three $\chi_{cj}$ states for studying hadronic 
transitions. The main hadronic transition process of $\chi_{c1}$ is $\chi_{c1}\to\eta_c\pi\pi$. In 
the framework of QCD multipole expansion, this is dominated by the E1-M1 transition, and the rate is 
significantly smaller than those of E1-E1 transitions such as $\psi'\to J/\psi~\pi\pi$. In addition, 
$\chi_{c1}$ cannot be directly produced in $e^+e^-$ collision. It can only be produced via the 
radiative transition $\psi'\to\gamma\chi_{c1}$. This is why $\chi_{c1}\to\eta_c\pi\pi$ is not easy
to detect at the original Beijing Spectrometer BES II. The designed luminosity of the upgraded Beijing 
Electron-Positron Collider BEPC II is about $1\--2$ orders of magnitude higher than that of the 
original BEPC I, and the ability of the upgraded Beijing Spectrometer BES III will be significantly 
improved compared with the original BES II, especially its photon detector. So that BES III will be 
able to detect processes like $\chi_{c1}\to\eta_c\pi\pi$ which BES II can hardly do. In this paper, we 
calculate the transition rate of $\chi_{c1}\to\eta_c\pi\pi$ in the framework of QCD multipole expansion, 
and our result shows that both the upgraded BES III and the Cornell CLEO-c have a good chance to study 
$\chi_{c1}\to\eta_c\pi\pi$. and the theoretical approaches to hadronic transitions can be further 
tested.

According to the formula for E1-M1 transitions \cite{KY81}, the transition amplitude for
$\chi_{c1}\to\eta_c\pi\pi$ is
\begin{eqnarray}                      
\displaystyle
&&\hspace{-0.4cm}{\cal M}_{E1M1}
=i\frac{g_Eg_M}{6m_c}\sum_{KL}\left(
\frac{\langle\eta_c|(s_c-s_{\bar{c}})_j|KL\rangle\langle KL|x_i|\chi_{c1}\rangle}
{E_I-E_{KL}}\right.\nonumber\\ 
&&\hspace{0cm}\left.+\frac{\langle\eta_c|x_i|KL\rangle\langle KL|(s_c-s_{\bar{c}})_j|\chi_{c1}\rangle}
{E_I-E_{KL}}\right)\langle\pi\pi|E^a_iB^a_j|0\rangle,
\label{E1M1}
\end{eqnarray}
where $\vec E^a$ and $\vec B^a$ are color electric and magnetic fields; $g_E$ and $g_M$ are
the corresponding coupling constants for E1 and M1 gluon emissions; 
$\vec x$ is the relative separation
between $c$ and $\bar c$ in the quarkonium; 
$s_c$ ($s_{\bar{c}}$) is the
spin of the $c$ ($\bar{c}$) quark; $E_I$ is the energy of the 
initial state $\chi_{c1}$; $E_{KL}$ and $|KL\rangle$ are the energy and eigenstate of the 
intermediate state with the principal quantum number $K$ and orbital angular momentum $L$, 
respectively. As described in Ref.~\cite{KY81}, we take the quark confining string model \cite{QCS} 
to describe the intermediate states.

The hadronization factor $\langle\pi\pi|E^a_iB^a_j|0\rangle$ is a second rank pseudo tensor at the
light hadron scale. In the rest frame of $\chi_{c1}$, it is a function of the two pion momenta
$q_1$ and $q_2$. Since the mass difference $M_{\chi_{c1}}-M_{\eta_c}=530.3$ MeV is small, the 
pions are soft. Taking the PCAC and soft pion approach, the hadronization factor is of the form
\begin{eqnarray}                         
\langle\pi\pi|E^a_iB^a_j|0\rangle=C\epsilon_{ijk}(q_1^kq_2^0+q_2^kq_1^0),
\label{hadronization}
\end{eqnarray}
where $C$ is approximately a constant characterizing the size of the matrix element. Unfortunately,
there is no experimental result serving as the input data to determine $C$ at present. Therefore 
we have to take an alternative approach to make predictions. Now we take the two gluon approximation
used in Refs.~\cite{KY81,KTY}, which is shown to be reasonable so far as only the transition rates 
are concerned \cite{KY81,KTY}. In this approach, we approximately express 
$\langle\pi\pi|E^a_iB^a_j|0\rangle$ as
\begin{eqnarray}                                  
\langle\pi\pi|E^a_iB^a_j|0\rangle\approx\langle\pi\pi|gg\rangle\langle gg|E^a_iB^a_j|0\rangle. 
\label{2gluons}
\end{eqnarray}
The matrix element $\langle\pi\pi|gg\rangle$ is at the scale of $M_{\chi_{c1}}-M_{\eta_c}\sim 530$ 
MeV, and is independent of the multipole gluon emission. If we take the same 
approximation to the process $\psi'\to J/\psi~\pi\pi$, i.e., expressing the corresponding hadronization 
factor as 
$\langle\pi\pi|E^a_iE^a_j|0\rangle\approx\langle\pi\pi|gg\rangle\langle gg|E^a_iE^a_j|0\rangle$, the
matrix element $\langle\pi\pi|gg\rangle$ in this process is at the scale of $M_{\psi'}-M_{J/\psi}
\sim 590$ MeV which is not much different from 530 MeV in the case of  $\chi_{c1}\to\eta_c\pi\pi$. 
Therefore we can neglect the running of $\langle\pi\pi|gg\rangle$ from 590 MeV to 530 MeV, and regard 
it as the same matrix element in both $\psi'\to J/\psi~\pi\pi$ and $\chi_{c1}\to\eta_c\pi\pi$. Then  
$\langle\pi\pi|gg\rangle$ can be absorbed into the definition of the phenomenological coupling 
constant $g_E$, and we can determine $g_E$ by taking the experimental datum of the transition rate
$\Gamma(\psi'\to J/\psi~\pi\pi)$ as input. The remaining factor $\langle gg|E^a_iB^a_j|0\rangle$
in (\ref{2gluons}) is then easy to evaluate. It has been shown in Refs.~\cite{KY81,KTY} that such an 
approach gives predictions for the transition rates quite close to those in the soft pion approach
\cite{footnote}.
So we take this approximation and obtain the following transition rate \cite{KY81}
\begin{eqnarray}                            
\Gamma(\chi_{c1}\to\eta_c\pi\pi)&=&\displaystyle\frac{4\alpha_E\alpha_M}{8505\pi m_c^2}
|f^{010}_{1110}+|f^{101}_{1110}|^2\nonumber\\
&&\times
\left(M_{\chi_{c1}}-M_{\eta_c}\right)^7,
\label{rate formula}
\end{eqnarray}
where $\alpha_E\equiv \displaystyle\frac{g_E^2}{4\pi}$, 
$\alpha_M\equiv \displaystyle\frac{g_M^2}{4\pi}$, and
\begin{widetext}
\begin{eqnarray}                                      
&&f^{LP_IP_F}_{n_Il_In_Fl_F}=\sum_K f^{LP_IP_F}_{n_Il_In_Fl_F}(K),\nonumber\\
&&f^{LP_IP_F}_{n_Il_In_Fl_F}(K)\equiv\displaystyle\frac{\int R^*_F(r'){r'}^{P_F}R_{KL}(r'){r'}^2dr'
\int R^*_{KL}(r){r}^{P_I}R_I(r){r}^2dr}{E_I-E_{KL}}
\label{f}
\end{eqnarray}
\end{widetext}
in which $R_I,~R_F,~R_{KL}$ are radial wave functions of the initial, final, and intermediate 
states, respectively. These radial wave functions can be obtained when a potential model for
the quarkonium is taken.

In this paper, we take the improved QCD motivated potential model proposed by Chen and Kuang 
(CK) \cite{CK}, and we take Potential I in Ref.~\cite{CK}. Comparing with other potential models, 
this potential model has the advantage that it reflects more about QCD with an explicit 
$\Lambda_{\overline{MS}}$ dependence, and leads to more successful phenomenological results
\cite{CK,Kuang06}.
The potential is
\begin{eqnarray}                      
\displaystyle
V(r)=kr-\frac{16\pi}{25}\frac{1}{rf(r)}\left[1+\frac{2\gamma_E+
\displaystyle\frac{53}{75}}{f(r)}-\frac{462}{625}\frac{\ln f(r)}{f(r)}\right],
\label{CKpot}
\end{eqnarray}
where $k=0.1491~{\rm GeV}^2$ is the string tension related to the Regge slope 
\cite{CK}, $\gamma_E$ is the Euler constant, and $f(r)$ is
\begin{widetext}
\begin{eqnarray}                      
\displaystyle
f(r)=\ln\left[\frac{1}{\Lambda_{\overline{MS}}~r}+4.62-\bigg(1-\frac{1}{4}
\frac{\Lambda_{\overline{MS}}}{\Lambda^I_{\overline{MS}}}\bigg)
\frac{1-\exp\bigg\{-\bigg[15\bigg(3\displaystyle
\frac{\Lambda^I_{\overline{MS}}}{\Lambda_{\overline{MS}}}
-1\bigg)\Lambda_{\overline{MS}}~r\bigg]^2\bigg\}}
{\Lambda_{\overline{MS}}~r}\right]^2,
\label{f(r)}
\end{eqnarray}
\end{widetext}
in which $\Lambda^I_{\overline{MS}}=180$ MeV. In this model, the $c$ quark mass is 
$m_c=1.478$ GeV, and we take $\Lambda_{\overline{MS}}=200$ MeV in the following calculation. 

With this potential model, we can calculate the transition amplitudes $f^{LP_IP_F}_{n_Il_In_Fl_F}(K)$
defined in Eq.~(\ref{f}). To see the convergence of the summation 
$\displaystyle\sum_K f^{LP_IP_F}_{n_Il_In_Fl_F}(K)$ in (\ref{f}), 
we list the values of the first eight amplitudes in TABLE I. We see that taking the first eight terms 
in the summation is enough. so we obtain
\begin{eqnarray}                      
f^{010}_{1110}+f^{101}_{1110}=-7.21402~{\rm GeV}^{-2}.
\label{fvalue}
\end{eqnarray}
\begin{widetext}
\begin{center}
\begin{table}[h]
\caption{The first eight transition amplitudes (in units of GeV$^{-2}$) obtained from the CK 
potential model.}
\tabcolsep 16pt
\begin{tabular}{cccc}
\hline\hline
&$f^{010}_{1110}(K)$ (GeV$^{-2}$)&$f^{101}_{1110}(K)$ (GeV$^{-2}$)
&$f^{010}_{1110}(K)+f^{101}_{1110}(K)$ (GeV$^{-2}$)\\
\hline
K=1&-2.43573&-5.14597&-7.58170\\
K=2&-0.17284&~0.50817&~0.33533\\
K=3&-0.02746&~0.05487&~0.02740\\
K=4&-0.00683&~0.01075&~0.00393\\
K=5&-0.00225&~0.00303&~0.00078\\
K=6&-0.00090&~0.00109&~0.00018\\
K=7&-0.00041&~0.00046&~~~4.5$\times 10^{-5}$\\
K=8&-0.00021&~0.00022&~~~9.0$\times 10^{-6}$\\
\hline
sum&-2.24676&-4.56726&-7.21402\\
\hline\hline
\end{tabular}
\end{table}
\end{center}
\end{widetext}

Next we consider the determination of $\alpha_E$ and $\alpha_M$. Taking the same approximation
to $\psi'\to J/\psi~\pi\pi$, we can obtain $\Gamma(\psi'\to J/\psi~\pi\pi)$ which is \cite{KY81}
\begin{eqnarray}                                
\Gamma(\psi'\to J/\psi~\pi\pi)=\frac{8\alpha_E^2}{8505\pi}|f^{111}_{2010}|^2(M_{\psi'}-M_{J/\psi})^7.
\label{2S-1Spipi}
\end{eqnarray}
The updated Particle Data Group (PDG) best fit experimental values \cite{PDG06}
\begin{eqnarray}                          
&&\Gamma(\psi')=337\pm 13~{\rm keV},\nonumber\\
&&B(\psi'\to J/\psi~\pi\pi)=(48.26\pm 0.95)\%,
\label{2S-1Sexpt}
\end{eqnarray}
leads to
\begin{eqnarray}                          
\Gamma(\psi'\to J/\psi~\pi\pi)=162.6\pm 9.4~{\rm keV}.
\label{2S-1Swidth}
\end{eqnarray}
$\alpha_E$ can then be determined by comparing the value of (\ref{2S-1Spipi}) with (\ref{2S-1Swidth}),
and this leads to
\begin{eqnarray}                          
\alpha_E=0.51.
\label{alpha_e}
\end{eqnarray}

The determination of $\alpha_M$ is not so direct. So far the only experimental result can be used as 
an input datum to determine $\alpha_M/\alpha_E$ is the CLEO-c experiment on searching for the $h_c$
state via $\psi'\to h_c+\pi^0$ \cite{CLEO-ch_c}. CLEO-c measured the product of the branching ratios
\cite{CLEO-ch_c}
\begin{widetext}
\begin{eqnarray}                         
B(\psi'\to h_c\pi^0)\times B(h_c\to\eta_c\gamma)=(3.5\pm1.0\pm0.7)\times 10^{-4}.
\label{CLEO-cexpt}
\end{eqnarray}
\end{widetext}
In an earlier paper \cite{Kuang02}, we calculated the related branching ratios by using QCD 
multipole expansion and the Gross-Treiman-Wilczek formula \cite{GTW} for obtaining the transition rate
$\Gamma(\psi'\to h_c\pi^0)$, and by using perturbative QCD for obtaining the hadronic decay width
$\Gamma(h_c\to hadrons)$. Our obtained result is \cite{Kuang02,Kuang06}
\begin{widetext}
\begin{eqnarray}                         
B(\psi'\to h_c\pi^0)\times B(h_c\to\eta_c\gamma)=1.94\left(\frac{\alpha_M}{\alpha_E}\right)
\times 10^{-4}.
\label{Kuangh_c}
\end{eqnarray}
\end{widetext}
Comparing these two results, we obtain
\begin{eqnarray}            
\frac{\alpha_M}{\alpha_E}=1.8\pm 0.9.
\label{M/E}
\end{eqnarray}
Note that there is a $50\%$ error in (\ref{M/E}) coming from the errors in (\ref{CLEO-cexpt}).

We would like to mention that the small pion mass effect \cite{NS} is neglected in 
Eq.~(\ref{rate formula}). 
That the pion mass effect is negligibly small was argued in Ref. \cite{Voloshin02}. In the 
present case, the uncertainty from the input value of $\alpha_M/\alpha_E$ in Eq.~(\ref{M/E}) 
is $50\%$. Compared with this large uncertainty, the small pion mass effect is 
evidently negilible in the present calculation.

With all these, we obtain the transition rate
\begin{eqnarray}                         
\null\hspace{-0.2cm}
\Gamma(\chi_{c1}\to\eta_c\pi\pi)=11.0\left(\frac{\alpha_M}{\alpha_E}\right){\rm keV}
=19.8\pm 9.9~{\rm keV}.
\label{rate}
\end{eqnarray}
In (\ref{rate}), the $50\%$ uncertainty comes from the error in (\ref{M/E}). For checking 
the model dependence of this prediction, we also calculated the transition rate using the well-known
Cornell model with Coulomb plus linear potential \cite{Cornell}, and the result is 
$\Gamma(\chi_{c1}\to\eta_c\pi\pi)=9.4\left(\alpha_M/\alpha_E\right)~{\rm keV}=17.0\pm8.5$ keV. 
So that the model dependence is about $14\%$ which is not significant considering the large
uncertainty in (\ref{rate}). The total width of $\chi_{c1}$ is 
$\Gamma(\chi_{c1})=0.89\pm 0.05$ MeV \cite{PDG06}. So that the branching ratio is
\begin{eqnarray}                             
B(\chi_{c1}\to\eta_c\pi\pi)=(2.22\pm1.24)\%.
\label{BR}
\end{eqnarray}

At $e^+e^-$ colliders, the state $\chi_{c1}$ can be produced via $\psi'\to\gamma\chi_{c1}$ at the 
$\psi'$ peak. In the rest frame of $\psi'$, the momentum of $\chi_{c1}$ is 171 MeV which is only $5\%$
of its mass $M_{\chi_{c1}}=3510.66$ MeV. Therefore we can neglect the motion of $\chi_{c1}$, and simply 
take the branching ratio (\ref{BR}) to estimate the event numbers in the experiments.

The detection of the process 
\begin{eqnarray}                           
\psi'\to\gamma\chi_{c1}\to\gamma\eta_c\pi\pi
\label{chain}
\end{eqnarray}
can be performed in two ways, namely the {\it inclusive} and the {\it exclusive} detections \cite{CLEO}.
In the inclusive detection, only the photon and the two pions are detected, while $\eta_c$ is regarded 
as a missing energy. In the exclusive detection, all the photon, the two pions, and the decay products 
of $\eta_c$ are detected. Reconstruction of $\eta_c$ and $\chi_{c1}$ from the measured final state 
tagging particles can suppress the backgrounds.

The inclusive detection requires measuring the momenta of the photon and the pions to certain 
precision. It is difficult to do this kind of analysis with the BES II data because the BES II photon 
detector is not efficient enough. At BES III and CLEO-c, this kind of detection is possible.

For BES III, it is not difficult to accumulate $10^8$ of $\psi'$ events. The branching ratio of
$\psi'\to\gamma\chi_{c1}$ is $B(\psi'\to\gamma\chi_{c1})=(8.7\pm0.4)\%$ \cite{PDG06}. Taking account of a 
$15\%$ detection efficiency, we obtain the number of events of (\ref{chain}) at BES III
\begin{eqnarray}                          
&&\null\hspace{-1cm}{\rm BES~III}:\nonumber\\
&&\null\hspace{-0.8cm}N_{incl}(\psi'\to\gamma\chi_{c1}\to\gamma\eta_c\pi\pi)=(2.90\pm1.74)\times 10^4.
\label{inclusiveBES III}
\end{eqnarray}
This is quite a large number.

CLEO-c is now running at the $\psi'$ peak again, and will soon accumulate $3\times 10^7$ $\psi'$ 
events. 
The CLEO-c detection efficiency is around $15\%$ \cite{CLEO-ch_c}. For such a sample , the 
corresponding number of events of (\ref{chain}) is
\begin{eqnarray}                          
&&\hspace{-1.2cm}{\rm CLEO-c}:\nonumber\\
&&\hspace{-1cm}N_{incl}(\psi'\to\gamma\chi_{c1}\to\gamma\eta_c\pi\pi)=(8.7\pm5.2)\times10^3.
\label{inclusiveCLEO-c}
\end{eqnarray}
So the transition (\ref{chain}) can also be clearly studied at CLEO-c.

For the exclusive detection, suitable decay modes of $\eta_c$ should be taken for identifying
the $\eta_c$ in (\ref{chain}). Some feasible decay modes with reasonable branching ratios are 
\cite{PDG06,CLEO-ch_c}
\begin{widetext}
\begin{eqnarray}                    
&&\eta_c\to\rho\rho:~~~~~~~~~~~~~~~~~~~~B(\eta_c\to\rho\rho)=(2.0\pm0.7)\%,
\label{rhorho}\\
&&\eta_c\to K^*\bar{K}^*:~~~~~~~~~~~~~~~B(\eta_c\to K^*\bar{K}^*)=(1.03\pm0.26)\%,
\label{KK}\\
&&\eta_c\to\phi\phi:~~~~~~~~~~~~~~~~~~~~B(\eta_c\to\phi\phi)=(0.27\pm0.09)\%,
\label{phiphi}\\
&&\eta_c\to K^*(892)^0K^-\pi^+:~~~B(\eta_c\to K^*(892)^0K^-\pi^+)=(2.0\pm0.7)\%,
\label{KKpi}\\
&&\eta_c\to K^+K^-\pi^+\pi^-:~~~~~~~B(\eta_c\to K^+K^-\pi^+\pi^-)=(1.5\pm 0.6)\%,
\label{KKpipi}\\
&&\eta_c\to 2(\pi^+\pi^-):~~~~~~~~~~~~~B(\eta_c\to 2(\pi^+\pi^-)=(1.20\pm0.30)\%,
\label{2pipi}\\
&&\eta_c\to\eta\pi\pi\to\gamma\gamma\pi\pi:~~~~~~~
B(\eta_c\to\eta\pi\pi\to\gamma\gamma\pi\pi)=(1.9\pm0.7)\%,
\label{etapipi}\\
&&\eta_c\to K_SK^\pm\pi^\mp:~~~~~~~~~~~B(\eta_c\to K_SK^\pm\pi^\mp)=(1.9\pm0.5)\%,
\label{K_SKpi}\\
&&\eta_c\to K_LK^\pm\pi^\mp:~~~~~~~~~~~B(\eta_c\to K_LK^\pm\pi^\mp)=(1.9\pm0.5)\%.
\label{K_LKpi}
\label{modes}
\end{eqnarray}
\end{widetext}
The modes (\ref{K_SKpi}) and (\ref{K_LKpi}) have been used by CLEO-c in the search 
for the $h_c$ state \cite{CLEO-ch_c}. 

 In the exclusive detection, the requirement of the precision of 
the photon momentum measurement in (\ref{chain}) is not so strict. So it is possible to do this kind
of analysis with the BES II data except for those modes with photons in $\eta_c$ decays. Therefore we 
also calculate the event numbers at BES II. The BES II data contains $1.4\times 10^7$ $\psi'$ events. 
Considering the ability of the BES II detector, we take a detection efficiency of $10\%$ for BES II. 
The Predicted event numbers for BES II, BES III, and CLEO-c are listed in TABLE II. We see that the 
exclusive detection of the process (\ref{chain}) can be well studied at BES III and CLEO-c. 
Considering the theoretical uncertainty, the events at BES II are marginal.
\begin{widetext}
\begin{center}
\begin{table}[h]
\caption{Predictions for the exclusive detection event numbers for the process (\ref{chain})
in the CK potential model using the
$\eta_c$ decay modes shown in (\ref{rhorho})$\--$(\ref{K_LKpi}) at
BES II, BES III, and CLEO-c. 
The accumulated numbers of the $\psi'$ events are taken to be $1.4\times 10^7$ for BES II, $10^8$
for BES III, and $3\times 10^7$ for CLEO-c. The detection efficiency is taken to be $10\%$ for 
BES II, and $15\%$ for BES III and CLEO-c.}
\tabcolsep 30pt
\null\vspace{0.1cm}
\begin{tabular}{cccc}
\hline\hline
modes&BES II&BES III&CLEO-c\\
\hline
$\rho\rho$&$54\pm51$&$579\pm552$&$174\pm165$\\
$K^*\bar{K}^*$&$28\pm24$&$298\pm255$&$90\pm76$\\
$\phi\phi$&7$\pm$7&$78\pm73$&$23\pm22$\\
$K^*(892)^0K^-\pi^+$&$54\pm51$&$579\pm552$&$174\pm165$\\
$K^+K^-\pi^+\pi^-$&$41\pm41$&$435\pm435$&$130\pm130$\\
$2(\pi^+\pi^-)$&$32\pm28$&$348\pm296$&$104\pm89$\\
$\eta\pi\pi\to\gamma\gamma\pi\pi$&$51\pm50$&$550\pm534$&$165\pm160$\\
$K_SK^\pm\pi^\mp$&$51\pm44$&$550\pm476$&$165\pm143$\\
$K_LK^\pm\pi^\mp$&$51\pm44$&$550\pm476$&$165\pm143$\\
\hline\hline
\end{tabular}
\end{table}
\end{center}
\end{widetext}

We have seen that the theoretical uncertainty mainly comes from the experimental errors in 
(\ref{CLEO-cexpt}). If the product of branching ratios 
$B(\psi'\to h_c\pi^0)\times B(h_c\to \eta_c\gamma)$ can be measured
to higher precision at CLEO-c or BES III, the predictions for the event numbers of the process 
(\ref{chain}) will be more definite.

Finally, we would like to mention that the numbers listed in TABLE II are obtained by taking 
the PDG best fit experimental value $\Gamma(\psi')=337\pm13$ keV in Eq.~(\ref{2S-1Sexpt}) as input. 
Actually, the Particle Data Book also provides a world averaged value $\Gamma(\psi')=277\pm 22$ keV
in Meson Particle Listing \cite{PDG06}. If we take this world averaged value as input instead of the PDG best fit value in 
(\ref{2S-1Sexpt}), all the numbers in TABLE II will be smaller by a factor of 0.82. This is 
another uncertainty caused by the input datum of $\Gamma(\psi')$. We expect further improved 
measurement of $\Gamma(\psi')$ to reduce this kind of uncertainty.

\null\noindent
{\bf Acknowledgement}

We would like to thank Roberto Mussa, Jia Liu, and Yuanning Gao for discussions. This work is supported
by the National Natural Science Foundation of China under the grant No.~90403017.



\begin{thebibliography}{100}
\bibitem{footnote}
Although there has been a comment on this approximation \cite{Voloshin86}, 
it has been discussed in Ref. \cite{KTY} why this approximation can lead to reasonable order
of magnitude estimates of the transition rates (except the case of very small phase space), and 
the predicted transition rates in this approximation are shown to be close to those obtained in 
the soft pion approach \cite{KY81}. Thus the two gluon approximation can be taken as a possible 
approximation for making order of magnitude estimate of transition rates when there is no 
reliable way of calculating the highly nonperturbative hadronization factor from QCD, and no 
suitable experimental input to determine the unknown coefficient in the soft pion formula 
(\ref{hadronization}). Considering the large uncertainty in the input datum (cf. Eq.~(\ref{M/E})
below), taking this approximation is reasonable in the present calculation.
\end{thebibliography}
\end{document}